\documentstyle{czjphys}         
\input{epsf}
  \def \bm #1{ \mbox {\boldmath $#1$} }

\begin{document}
\title{ANGLE DEPENDENCE OF POLARIZATION OBSERVABLES
       BY FRAGMENTATION OF DEUTERONS TO PIONS}        
\authori{\underline{A.Yu. Illarionov}, A.G. Litvinenko, G.I. Lykasov}
\addressi{Joint Institute for Nuclear Research,\\
          141980 Dubna, Moscow Region, Russia}     
\authorii{}     
\addressii{}    
\authoriii{}    
\addressiii{}   
\headtitle{ANGLE DEPENDENCE OF POLARIZATION OBSERVABLES
           BY FRAGMENTATION \ldots } 
\headauthor{\underline{A.Yu. Illarionov}, A.G. Litvinenko, G.I. Lykasov}
\specialhead{\underline{A.Yu. Illarionov}, A.G. Litvinenko, G.I. Lykasov:
 ANGLE DEPENDENCE OF POLARIZATION OBSERVABLES \ldots }
\evidence{A}
\daterec{XXX}    
\cislo{0}  \year{2001}
\setcounter{page}{1}
\pagesfromto{000--000}
\maketitle

\begin{abstract}
  The fragmentation of deuterons into pions with nonzero angle emitted in the
  kinematical region forbidden for free nucleon-nucleon collisions
  is analyzed. The inclusive relativistic invariant spectrum of pions and
  the tensor analyzing power ${\rm A}_{YY}$ are investigated within the
  framework of an impulse approximation using different kinds of the deuteron
  wave function. The influence of ${\rm P}$-wave contribution to the
  deuteron wave function is studied, too. Our results are compared with the
  experimental data and other calculations performed within both the
  non-relativistic and relativistic approaches. It was found that
  theoretical calculations based on IA do not provide a consistent
  understanding of the new data in a whole cumulative region that can be
  caused by a influence of non-nucleon degrees of freedom.
\end{abstract}

 \section{Introduction}
 \label{sec:intro}

 Recently the intriguing preliminary data obtain at ``SPHERE''
group (LHE JINR) on the deuteron tensor analyzing power
${\rm A}_{YY}$, measured in the reaction on fragmentation of tensor
polarized deuterons into cumulative pions with nonzero pion production
angle, have been reported \cite{zol98,lit00}.
They are rather precise and correspond to a maximum cumulative
variable ${\rm x_C}$ of $1.75$ (internal momentum up to $500$ MeV/c).
Since in the $pD$ collisions the deuteron gets from proton a momentum
comparable with the deuteron mass, these data are probing the truly
relativistic dynamics inside the deuteron \cite{fra81}.

The interest of deuterons fragmentation into pions arises from the
possibility
 (I) to experimental study of the polarization observables of
deuteron fragmentation into hadrons containing of different quarks,
additionally to well-famous proton stripping reaction
\cite{lyk93,abl83,abl90},
 (II) to theoretical study of nuclear structure at short distances on a based
of available deuteron models,
 (III) to develop in the future the reasonable theory of the relativistic
nuclear systems.
The measurement of the unpolarized characteristics together with polarization
observables allows us to analyze this processes more correctly and to
conclude that together with the scheme of the deuteron wave function (DWF)
relativization the relativistic description of such processes themselves is
of important as well.

 In this talk we present a relativistic invariant analysis of the deuteron
tensor analyzing power ${\rm A}_{YY}$ performed in the framework of the
relativistic impulse approximation at nonzero pion production angle
(see, for review, the paper \cite{ill00}, which contains also our calculation
of ${\rm T}_{20} = - \sqrt{2} {\rm A}_{YY}$ at zero pion production angle).

 \section{Relativistic impulse approximation}
 \label{sec:RIA}

\begin{figure} \centering
\mbox{\epsfysize=4cm \epsffile{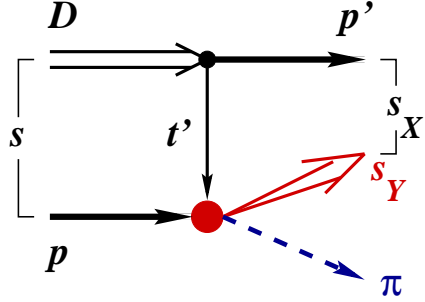}}
\caption{
 \label{fig1} Impulse approximation diagram.
}
\end{figure}

 Let us consider the inclusive reaction of fragmentation of tensor
polarized deuteron to pion, ${\bm D}(p,\pi)X$, where the typical initial
energy of a few GeV, and the final pion is detected with nonzero
angle $\theta_\pi$. Within spectator mechanism approach this reaction
presented by the impulse approximation diagram shown in figure \ref{fig1},
where the upper and lower vertices should factorize \cite{ill00} and
consequently they may be computed separately.

\noindent
 If the initial deuteron is only aligned due to $p_D^{YY}$ component, then
 the inclusive spectrum of this reaction can be written in the form:
 \begin{eqnarray}
 \rho_{pD}^\pi\left(p_D^{YY}\right)=
 \rho_{pD}^\pi \Bigl[1 + {\rm A}_{YY} \cdot p_D^{YY} \Bigl]~,
 \label{def:AYY}
 \end{eqnarray}
 where
 $\rho_{pD}^\pi \equiv \varepsilon_\pi \cdot {d\sigma_{pD}^\pi / d^3p_\pi}$
 is the inclusive spectrum for the case of unpolarized
 deuterons and ${\rm A}_{YY}$ is the tensor analyzing power. They can be
 written in a fully covariant manner within the Bethe-Salpeter formalism.
 This way is possible to come to general conclusions about the amplitude of
 the process not to seen in the non-relativistic approach. Such analysis
 of the deuteron models is very important assuming the searching of nuclear
 quarks phenomena \cite{wal00}. One can write the observables
 (\ref{def:AYY}) in the factorization form \cite{ill00}:
 \begin{eqnarray}
 &&\rho_{pD}^\pi = 
 {1\over(2\pi)^3} \int {\sqrt{\lambda(p,n)}\over\sqrt{\lambda(p,D)}}~\left[
 \rho_{pN}^\pi\cdot\Phi^{(u)}(|\bm q|)\right]~{d^3q\over E_{\bm q}}~;
 \label{un.cs} \\
 &&\rho_{pD}^\pi\cdot{\rm A}_{YY} =
\nonumber \\ &&\hspace{0.6cm}
 {1\over(2\pi)^3} \int {\sqrt{\lambda(p,n)}\over\sqrt{\lambda(p,D)}}~\left[
 \rho_{pN}^\pi\cdot\Phi^{(t)}(|\bm q|)\right]
 \left({1 - 3 \sin^2\theta_{\bm q} \sin^2\varphi_{\bm q} \over 2}\right)
~{d^3q\over E_{\bm q}}
 \label{calc:AYY}
 \end{eqnarray}
 where $\lambda(p_1, p_2) \equiv (p_1p_2)^2 - m_1^2 m_2^2 =
 \lambda(s_{12}, m_1^2, m_2^2) / 4$ is the flux factor; $p, n$ are
 the four-momenta of the proton-target and intra-deuteron nucleon,
 respectively;
 $\rho_{pN}^\pi \equiv \varepsilon_\pi \cdot {d\sigma_{pN}^\pi / d^3p_\pi}$
 is the relativistic invariant inclusive spectrum of pions produced by
 interacting the intra-deuteron nucleon with the proton-target.
 In the general case, this
 spectrum can be written as a three-variable function
 $\rho_{pN}^{\pi} = \rho(x_f, \pi_\perp, s_{NN})$. Feynman's variable,
 $x_f$, is defined as $x_f = 2\pi_{||} / \sqrt{s_{NN}}$,
 where $\pi$ is the pion momentum in the center of mass of two
 interacting nucleons and $s_{NN} = (p + n)^2$. As shown in \cite{ill00},
 ${\rm A}_{YY}$ is more sensitive to the DWF form than this invariant
 spectrum.

 The functions $\Phi^{(u)}(|\bm q|)$ and  $\Phi^{(t)}(|\bm q|)$ depends on
 the relative momentum $q = (n - p')/2$ and contains full information
 about the structure of deuteron with one on-shell nucleon.
 They can be written in terms of positive-energy, $U=^3{\cal S}^{++}_1,~
 W=^3{\cal D}^{++}_1$, and negative-energy, $V_s=^1{\cal P}^{-+}_1$,~
 $V_t=^3{\cal P}^{-+}_1$, wave functions:
 \begin{eqnarray}
 \Phi^{(u)}(|\bm q|) &=& U^2 + W^2 - V_t^2 - V_s^2
\nonumber \\
 &+& {2\over\sqrt3}{|\bm q|\over m} \left[
 \left(\sqrt2V_t - V_s\right)U - \left(V_t + \sqrt2 V_s\right)W\right]~;
 \label{Phi(u)} \\
 \Phi^{(t)}(|\bm q|) &=& 2\sqrt2 UW + W^2 + V_t^2 - 2V_s^2
\nonumber \\
 &-& {4\over\sqrt3}{|\bm q|\over m}\left[\left(U - {W \over \sqrt2}\right)
  {V_t \over \sqrt2} + \left(U + \sqrt2 W\right)V_s\right]~.
 \label{Phi(t)}
 \end{eqnarray}

 It is intuitively clear that the two nucleons in the deuteron are mainly
 in states with angular momentum $L = 0, 2$ (see also numerical analysis of
 the solutions of the $BS$ equation in terms of amplitudes within the
 $\rho$-spin basis \cite{kap96}), so the probability of states with $L = 1$
 $(V_{s,t})$ in Eqs.(\ref{Phi(u)},\ref{Phi(t)}) is much smaller in
 comparison with the probability for the $U, W$ configurations. Moreover,
 it can be shown that the $U$ and $W$ waves directly correspond to the
 non-relativistic ${\rm S}$ and ${\rm D}$ ones. Therefore,
 Eqs.(\ref{Phi(u)},\ref{Phi(t)}) with only $U, W$ waves can be identified
 as the main contributions to the corresponding
 observables and they may be compared with their non-relativistic analogies.
 The other terms posses contributions from the ${\rm P}$-waves and they are
 proportional to $\bm q/m$ (the diagonal terms in $V _{s,t}$ are negligible).
 Due to their pure relativistic origin one can refer to them as relativistic
 corrections.

 The natural way to compare the non-relativistic and relativistic
 calculations is the Kamada-Gl$\ddot{\mbox{o}}$ckle method \cite{kam98}
 that is used to make relativistic quantum models using a realistic
 nonrelativistic nucleon-nucleon interaction as input, which requires the
 operator for the kinetic energy is formed out of square roots. This
 defines the scale transformation between the relativistic momenta $\bm q$
 in Eqs.(\ref{Phi(u)},\ref{Phi(t)}) and nonrelativistic momenta $\bm q_{KG}$:
\begin{equation}
 2\sqrt{\bm q^2 + m^2} := 2m + {\bm q_{KG}^2 \over 2m}
~~\to~~
 \bm q_{KG}^2 = 2m (E_{\bm q} - m)~,
\label{def:qKG}
\end{equation}
and renormalization of the DWF, that simply ensures that
the change of variables is unitary:
\begin{equation}
 \Psi_D(\bm q) = {\Psi_{N.R.}[\bm q_{KG}(\bm q)] \over h[\bm q_{KG}(\bm q)]}~,
\label{DWF-KG}
\end{equation}
where we defined the function $h(q)$ by the equation:
\begin{equation}
 q^2dq = h^2(q_{KG})q_{KG}^2dq_{KG}
~~\to~~
 h^2(q_{KG}) = {q \over q_{KG}}\left(1 + {q_{KG}^2 \over 2m^2}\right)~.
\label{H-KG}
\end{equation}
Then, the transformation involves the simple rescaling and the
renormalization, that ensures unitarity of an ${\rm S}$-matrix. The minima
for deuteron components are shifted towards larger momenta. The effects are
large above about $5$ fm$^{-1}$ ($\bm q^2/4m^2 \sim 0.25$).

 Let us consider the ``minimal relativization scheme'' describes rather well
 the differential cross section for such process as deuteron break-up
 $A(D,p)X$. The minimal relativization procedure \cite{bro73,fra81} consist
 of (i) a replacement of the argument of the non-relativistic wave functions
 by a light-cone variable $\bm k = (\bm k_\perp,\bm k_{||})$
 \begin{eqnarray}
 \bm k^2={m^2 + \bm k_\perp^2 \over 4x(1-x)}-m^2~;~~
 k_{||}=\sqrt{m^2 + \bm k_\perp^2 \over x(1-x)}\left({1\over2} - x\right)~.
 \label{def:lcv}
 \end{eqnarray}
 where
 $x=(E_{\bm q} + |\bm q|\cos\theta_{\bm q})/M = (\varepsilon'-p'_{||})/M~;~~
 |\bm k_\perp| = p'_\perp$ in the deuteron rest frame, and (ii) multiplying
 the wave functions by the factor $\sim$ $1/(1-x)$. As a results the
 argument is shifted towards smaller values and the wave function itself
 decreases less rapidly. This effect of increasing the wave function is
 compensated by the kinematical factor $1/(1-x)$.

 \section{Results and discussion}
 \label{sec:res}

 The results of calculation are summarized in figures below, where the
tensor analyzing power ${\rm A}_{YY}$ as function of the cumulative
${\rm x_C}$ (``cumulative number'' \cite{sta79}) is shown. For our reaction,
this variable is defined as follows:
\begin{eqnarray}
 {\rm x_C} = 2 {(p\pi) - \mu^2/2 \over (Dp) - Mm - (D\pi)} \leq 2~.
\label{xC}
\end{eqnarray}
The value of ${\rm x_C}$ corresponds to a minimum mass (in nucleon mass
units) of part of the projectile nucleus (deuteron) involved in the
reaction. When the pion with ${\rm x_C}$ larger than 1 is produced, it is
assigned to the cumulative pion. This kinematical region corresponds to the
values of the light-cone variable $x \ge 1$ (\ref{def:lcv}) and internal
deuteron momentum $ \ge 0$, as presented in the following table
for the planed case \cite{lit00} of pion production angle
$\theta_\pi = 178$ mrad.

\begin{center}
\begin{tabular}{||c||c|c|c|c|c|c|c|c||}
 \hline \hline
${\rm x_C}$&0.0 -- 1.0&1.2&1.4&1.5&1.6&1.7&1.8&1.9\\
 \hline
$|\bm q|_{min}~ \mbox{(GeV)}$&0.0&0.14&0.29&0.39&0.48&0.61&0.75&0.96\\
 \hline
$|\bm q_{KG}|_{min}~ \mbox{(GeV)}$&0.0&0.14&0.29&0.38&0.47&0.58&0.70&0.87\\
 \hline
$|\bm k|_{min}~ \mbox{(GeV)}$&0.0&0.13&0.25&0.33&0.40&0.48&0.57&0.69\\
 \hline \hline
\end{tabular}\\[0.1cm]
\end{center}

Deuteron fragmentation into proton $D + A \to p(0^o) + X$ is one of the
more intensively studied reaction with hadronic probe.  As shown in
\cite{pun96}, both the differential cross section and ${\rm T}_{20}$ for
fragmentation $D + p \to p(0^o) + X$ can be described within the IA
up to $k \leq 0.2$~GeV/c only. The inclusion of correction to
IA related to secondary interactions allows one to describe the
experimental data on the deuteron fragmentation $D p \to p X$ at
$k > 0.25$~GeV/c \cite{lyk93}.

\begin{figure} \centering
\mbox{\epsfysize = 8cm \epsffile{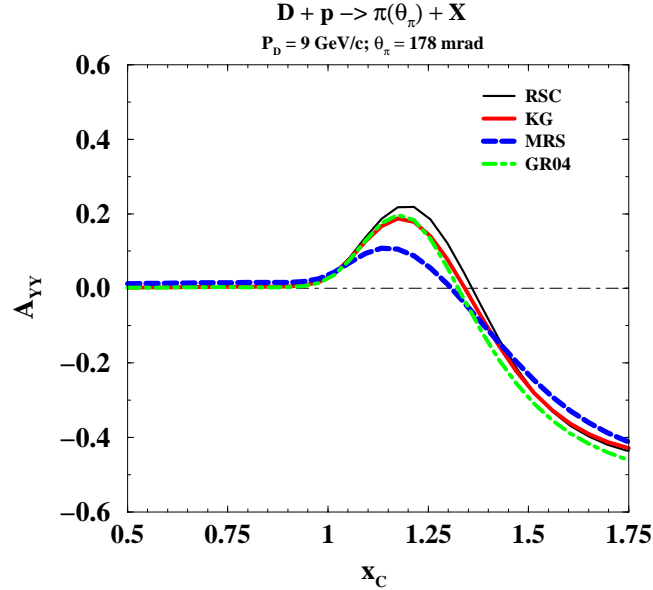}}
\caption{
 The tensor analyzing power ${\rm A}_{YY}$ for the planed \cite{lit00} pion
production angle $\theta_\pi = 178$ mrad, calculated using the
various types of the DWF. The thin solid line corresponds to the calculus
with the Reid DWF \cite{rei68}. The dot-dashed line represent the calculation
with the Gross DWF \cite{gro79} with the ${\rm P}$-wave probability
$P_V = \int_0^\infty q^2dq \cdot [V_t^2 + V_s^2] \approx 0.44\%$.
The thick solid line and long-dashed line calculated with the 
Reid DWF using the Kamada-Gl$\ddot{\mbox{o}}$ckle (KG) method \cite{kam98}
and the minimal relativization scheme (MRS) \cite{fra81,bro73}.
 \label{fig2} 
}
\end{figure}

\begin{figure} \centering
\mbox{\epsfysize = 8cm \epsffile{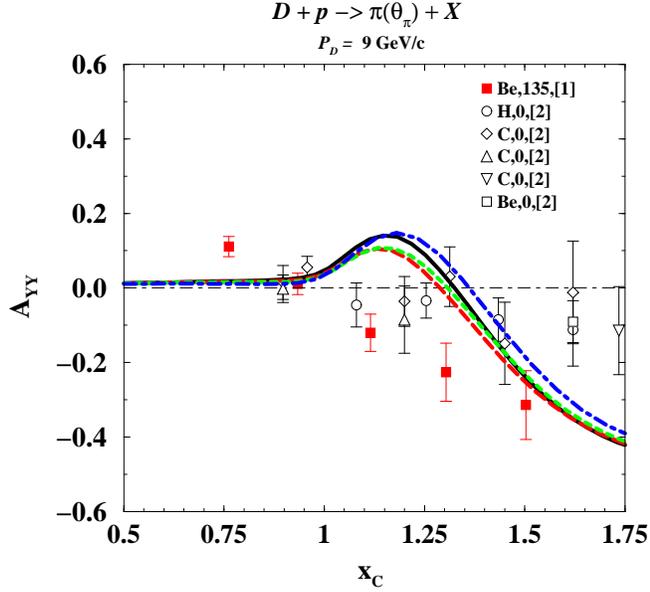}}
\caption{
 The angle dependence of the tensor analyzing power ${\rm A}_{YY}$
of the deuterons, calculated using the minimal relativization scheme
(MRS) \cite{fra81,bro73} for the projectile deuteron momentum
$P_D = 9$ GeV/c. The solid, long-dashed, dashed and dot-dashed lines
correspond to the calculus at $\theta_\pi = 0, 135, 178$ and $300$
mrad, respectively.
The  experimental data are taken from \cite{zol98,lit00,afa98}.
 \label{fig3} 
}
\end{figure}

The calculated results of ${\rm A}_{YY}$ for the reaction of polarized
deuteron fragmentation into cumulative pions are shown in figures \ref{fig2}
and \ref{fig3}. From this figures one can see that it is quite incorrect to
use the nonrelativistic DWF for the analysis of deuteron fragmentation into
pions. Relativistic effects are sizable, especially in the kinematic region
corresponding to short intra-deuteron distances or large ${\rm x_C}$.

The figure \ref{fig3} shows a small dependence of the tensor analyzing power
${\rm A}_{YY}$ to the pion production angle on the contrary to the
experimental data. On the other hand, the experimental data on ${\rm A}_{YY}$
are not described by any DWF used in this calculus over all region
${\rm x_C} \ge 1$.

 \section{Questions}
 \label{sec:quest}

  For a present time, the most sophisticated theoretical approaches are
unable to describe properly the experimental data in the cumulative
region of this notoriously difficult reaction. Then the following questions
still open:

\begin{enumerate}
 \item
 Why the measured ${\rm T}_{20}$ and ${\rm A}_{YY}$ for deuteron
fragmentation into pions are not reproduce by IA even for internal
momentum $< 250$ GeV/c where the same observables for deuteron breakup
reaction are in a good agreement with IA?

$\bullet$~~Additional to IA mechanisms \cite{kon75}?

\item
 Moreover the data on ${\rm T}_{20}$ at zero angle have small value as it
have to be for case of isotropic source from which pions are emitted.
The calculations presented in this work performed in the framework 
of nucleon model of deuteron. For the small inter-nucleon distances
(equal or less of hadron size) the use of nucleon as a quasi-particle
seams to be groundless and effects of manifestation of non-nucleon
degrees of freedom in deuteron could be expected.
 In spite of this the data on ${\rm A}_{YY}$ can be cause by

$\bullet$~~non-nucleonic degrees of freedom \cite{lyk93,glo93,kob93}.
\end{enumerate}


\end {document}